\documentclass[conference]{IEEEtran}
\usepackage[pdftex]{graphicx}
\usepackage{graphicx}
\usepackage[dvips]{epsfig}
\usepackage[caption=false,font=footnotesize]{subfig}

\usepackage{amsmath}

\usepackage{cite}

\usepackage{url}

\usepackage{siunitx}

\usepackage{lipsum}

\usepackage[mathscr]{euscript}
\usepackage{booktabs}
\usepackage{comment}
\usepackage[flushleft]{threeparttable} % For tablenotes

\usepackage{xcolor,colortbl}
\usepackage{color,soul}

\definecolor{morange}{rgb}{0.8,0.2,0}
\definecolor{mblue}{rgb}{0,0.3,1.0}
\definecolor{mpink}{rgb}{1.0,0.6,0.6}
\definecolor{mgreen}{rgb}{0.1,0.6,0.2}
\definecolor{mgoodgreen}{rgb}{0.9,1.0,0.7}

\definecolor{Gray}{gray}{0.85}

\usepackage{multirow}
\usepackage{array}
\newcolumntype{L}[1]{>{\raggedright\let\newline\\\arraybackslash\hspace{0pt}}m{#1}}
\newcolumntype{C}[1]{>{\centering\let\newline\\\arraybackslash\hspace{0pt}}m{#1}}
\newcolumntype{R}[1]{>{\raggedleft\let\newline\\\arraybackslash\hspace{0pt}}m{#1}}

\newcolumntype{G}{>{\columncolor{mgoodgreen}}c}

\newcommand{\dtoprule}{\specialrule{1pt}{0pt}{0.4pt}%
            \specialrule{0.3pt}{0pt}{\belowrulesep}%
            }
\newcommand{\dbottomrule}{\specialrule{0.3pt}{0pt}{0.4pt}%
            \specialrule{1pt}{0pt}{\belowrulesep}%
            }

\begin{document}

 \title{Demo: MmWave Lens MIMO}

% author names and affiliations
% use a multiple column layout for up to three different
% affiliations
%\author{\IEEEauthorblockN{Changmin Lee*, H. Birkan Yilmaz*, Chan-Byoung Chae*, Nariman Farsad**, and Andrea Goldsmith**}
%\IEEEauthorblockA{*School of Integrated Technology,  Institute of Convergence Technology, Yonsei University, Korea \\
%**Department of Electrical Engineering, Stanford University, USA \\
%Email:\{cm.lee, birkan.yilmaz, cbchae\}@yonsei.ac.kr, \{nfarsad, andrea\}@stanford.edu}%
%}

\begin{comment}
\author{
\IEEEauthorblockN{Sang-Hyun Park, Dongsoo Jun, Dong Ku Kim \\and Chan-Byoung Chae}
\IEEEauthorblockA{School of Integrated Technology, Yonsei University, Korea\\ 
Email:\{williampark, dongsoo.jun, dkkim, cbchae\}@yonsei.ac.kr}
\and
\IEEEauthorblockN{Byoungnam Kim and Inseok Jang}

\\
\IEEEauthorblockA{SensorView, Korea\\
Email: \{klaus.kim, bob.jang\}@sensor-view.com}
}
\end{comment}

\begin{comment}
\author{
\IEEEauthorblockN{Sang-Hyun Park, Dongsoo Jun,  Inseok Jang, Byoungnam Kim, Dong Ku Kim and Chan-Byoung Chae}
\IEEEauthorblockA{School of Integrated Technology, Yonsei University, Korea\\
Sensor View, Korea\\
Email:\{williampark, dongsoo.jun\}@yonsei.ac.kr, \{klaus.kim, bob.jang\}@sensor-view.com, \{dkkim, cbchae\}@yonsei.ac.kr}
}
\end{comment}

\author{
\IEEEauthorblockN{Sang-Hyun Park, Dongsoo Jun, Byoungnam Kim, Dong Ku Kim and Chan-Byoung Chae}
\IEEEauthorblockA{School of Integrated Technology, Yonsei University, Korea\\
Sensor View, Korea\\
Email:\{williampark, dongsoo.jun, dkkim, cbchae\}@yonsei.ac.kr, klaus.kim@sensor-view.com}
}

% conference papers do not typically use \thanks and this command
% is locked out in conference mode. If really needed, such as for
% the acknowledgment of grants, issue a \IEEEoverridecommandlockouts
% after \documentclass

\vspace{-10pt}
% make the title area
\maketitle
\begin{abstract}

%5G MIMO 시스템에서, analog digital hybrid 에서 high antenna gain 과 low hardware cost의 장점으로 렌즈안테나가 소개되었다. 이 페이퍼에서는 우리는 시뮬레이션을 통한 mmWave 통신에 가장 적합한 렌즈를 디자인하고, 제작하여, 리얼타임  mmWave lens MIMO testbed를 선보인다. 또한 기존 phased shift MIMO를 사용하는 통신환경에서,  mmwave의 wideband 상황에서 도래되는 문제인 beam-squint 현상을 제작한 lens 안테나로 평가한다. 

In fifth-generation (5G) communication, the RF lens antenna is introduced with advantages of high antenna gain and low hardware cost in analog/digital hybrid beamforming technology. In this paper, we design and manufacture the appropriate RF lens antenna for mmWave communication identified in simulation analysis and present a real-time mmWave lens MIMO testbed. Furthermore, in conventional phase-shift MIMO, beam-squint, a problem caused by the wideband of mmWave communication, is estimated with the fabricated RF lens antenna.

%The proposed technique uses machine-learning techniques and may be utilized in other studies that assume two point transmitters and two spherical absorbing receivers.
% It is important then to model the molecular MIMO channel; the modeling precess, though, is complicated by the existence of two antennas. Existing studies related to channel modeling in molecular single-input single-output (SISO) systems are analytically done for some simple environments. In contrast, molecular MIMO communication systems have no exact channel model. In this paper, we introduce a technique for modeling the molecular MIMO channel with diverse environments. The proposed technique uses machine-learning techniques and may be utilized in other studies that assume two point transmitters and two spherical absorbing receivers.
\end{abstract}

% no keywords

\IEEEpeerreviewmaketitle

\vspace{-5pt}

% % % % % % % % % % % % % % % % % % % % % % % % % % % % % % %
% % % % % % % % % % % % % % % % % % % % % % % % % % % % % % %
% % % % % % % % % % % % % % % % % % % % % % % % % % % % % % %
\section{Introduction}
 With the realization of 5G wireless communications requiring wide bandwidth and high spectral efficiency, the understanding of the millimeter wavelengths ~(6-300~GHz) of electromagnetic waves and the advanced multiple-input and multiple-output (MIMO) technology has become essential. In the mmWave MIMO system, a lens antenna array structure incorporating an RF lens is proposed to achieve high antenna gain with low hardware cost  \cite{sayeed2011continuous}. The narrow beamwidth of the lens antenna can preserve the reduced RF chain operation, making it possible to reduce the power consumption and interference between the streams.  
	%Recent researches in~\cite{sayeed2011continuous} proposed a convex lens system, transferring the signals towards different points of the focal surface. 
	
	High antenna gain and directivity can be achieved using the lens due to its properties, but challenges such as the low beam scanning angle exist because of the physical limitations of the lens. Therefore, we analyze the issues related to lens antenna design to obtain the desired performance in a realistic environment. Then, we consider the physical characteristics, such as lens structure, shape, material, and $f/D$ ratio, and carry out an analysis on the effect of the lens antenna design on the communication performance. As a result, we describe the relationship between the characteristics and performance of the lens antenna and find the suitable design conditions for desired communication performance. 
	%In addition, few studies have analyzed the effect of optical and electromagnetic parameters on lens antenna system performance. 
	
	In addition, when we use a wide frequency band, conventional analog beamforming is significantly affected by the beam-squint phenomenon. In a phased array antenna, the beam-squint means that the transmission angle and direction of the analog beamforming of the antenna are changed depending on the operating frequency \cite{DefBQ}. A hybrid beamforming system combining a lens antenna has been proposed to reduce the system cost, but the beam-squint problem in ultra-wideband systems has not been considered extensively. Therefore, in the case of wideband mmWave communication, it is essential to analyze the beam-squint of the RF lens and compare it with the phased array antenna system in the wideband.
	% In the first approach considering the beam-squint effect on lens embedded mmWave system, we also suggest an additoinal benefit of a lens antenna. When we consider the lens antenna design and property, well designed lens can make robustness to beam-squint problem.
		
	Then, the actual lens antenna is fabricated and a link-level experiment is demonstrated to show the results and verify our analysis. Finally, we analyze the overall performance in a realistic communication environment using the software-defined radio (SDR) platform. As a result, this paper demonstrates that the lens antenna design and beam-squint problem affect communication performance in an mmWave MIMO system.

% % % % % % % % % % % % % % % % % % % % % % % % % % % % % % %
% % % % % % % % % % % % % % % % % % % % % % % % % % % % % % %
% % % % % % % % % % % % % % % % % % % % % % % % % % % % % % %

%\begin{figure}[t!]
%	\centering{\includegraphics[width=1\columnwidth,keepaspectratio]{figures/Lens_structure.eps}}
%	\caption{(a) Lens antenna communication system model (b) Hyperbolic lens shape formulated in ractangular coordinates (c) Properties of a lens} \label{fig3:SULA}
%	\end{figure}

\vspace{-5pt}

\section{System Model}
\subsection{Hardware Layout} 
% % % %
% 이 통신 시스템은 랩뷰 시스템 디자인 소프트웨어 PXIe software-defined radio로 이루어져 있다. 그림 2번에서 보듯이, Tx 와 Rx가 각 셰시로 존재하여 떨어져 있다. Tx 에는 Chassis에 IF-LO module, I/Q-generator, Tx upconveter가 각각 NI PXIe-3620, NI PXIe-3610, NI mmRH-3642 로 이루어 져 있고, Rx는 IF-LO module, I/Q-digitizer, Decoder, Multiple access channel (MAC), Rx downconverter가 NI PXIe-3620, PXIE-3630, PXIe-7976R, PXIe 7976R, MMrh-3652로 이루어 져 있다. Tx 와 Rx의 안테나는 모두 Lens antenna 로 구현 되고 이 렌즈안테나는 고려되는 파라미터를 만족하게끔 제작되었으며, 백플레인에는 analog beamforming을 위한 2by2 patch antenna가 존재한다. 

The communication system consists of the LabVIEW system design software and PXIe (SDR). As shown in Fig. \ref{Fig.Testbed}, the Tx node as a base station and the Rx node as a user are separated by 5 meters. The Tx consists of the IF-LO module, I/Q-generator, and Tx NI upconverter as NI PXIe-3620, NI PXIe- 3610, and NI mmRH-3642, respectively. The Rx consists of the IF-LO module, I/Q-digitizer, decoder, multiple-access channel (MAC), and Rx downconverter as NI PXIe-3620, PXIE-3630, PXIe-7976R, PXIe 7976R, and mmRH-3652, respectively. The fabricated lens antenna is used in both the Tx and Rx. In the backplane of the RF lens antenna, a 2 by 2 patch antenna exists. 

%%%%%%%%%%%%%%%%%%%%%%%%%%%%%%%%%%%%%%%%%%

\subsection{Lens Antenna and Beam-squint} 
% Lens antenna는 PASSIVE BEAMFORMING 장비이기에 제작에 고려해야할 파라미터들이 있다. 첫번째로 As the material having a high permittivity, the beam shape performance deteriorated instead of increasing the beam scanning angle.  두번째로  if we want to increase the signal resolution to improve signal processing performance in multi-user scenarios, designing a lens antenna with a high $f/D$ ratio should be considered. In this demonstration, the lens shape is designed by applying a hyperbolic lens, and the lens antenna can be fabricated using polyethylene with $f/D$ ratio 0.7. Experiments with this lens antenna system resulted in beam gains of up to 21.8dB.

\begin{figure}[h]
	\centerline{\resizebox{1\columnwidth}{!}{\includegraphics{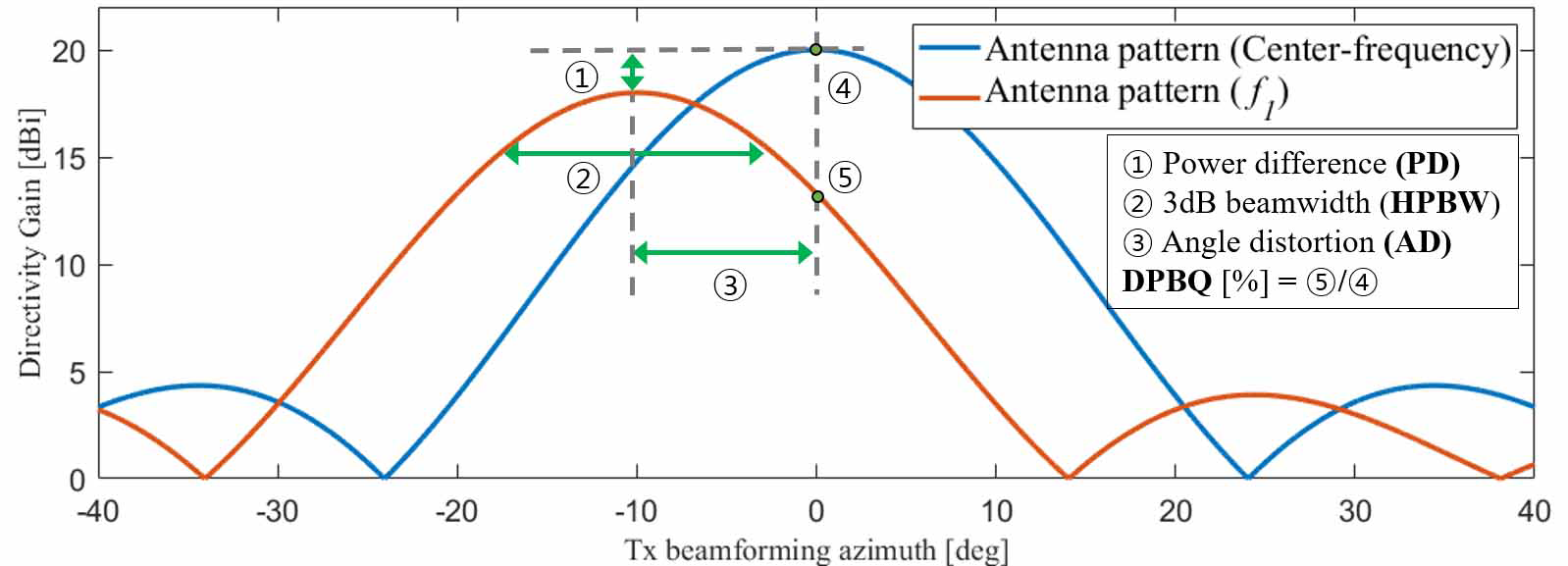}}}
	\caption{Relation between PD, AD, HPBW and DPBQ}\vspace{-1pt}
	\label{Fig.Relation}
\end{figure}

There are parameters to consider in the manufacture of the RF lens antenna, which is passive beamforming equipment. As the material has high permittivity, high permittivity causes high propagation loss in the lens that decreases the beam gain. Because scanning angle and beam gain are trade-off relation over permittivity of a material, using a material having a high permittivity to improve scanning angle causes severe beam gain degradation. If we want to increase the scanning angle in a fixed material, designing a lens antenna with a low $f/D$ ratio should be considered. In this demonstration, the lens shape is designed by applying a hyperbolic lens, and the lens antenna can be fabricated using polyethylene with a $f/D$ ratio of 0.7. Experiments with this lens antenna system resulted in beam gains of up to 21.8 dB.

\begin{figure}[t!]
\centering
\subfloat[\label{Fig.link_fig}]{{\resizebox{0.699\columnwidth}{!}{\includegraphics{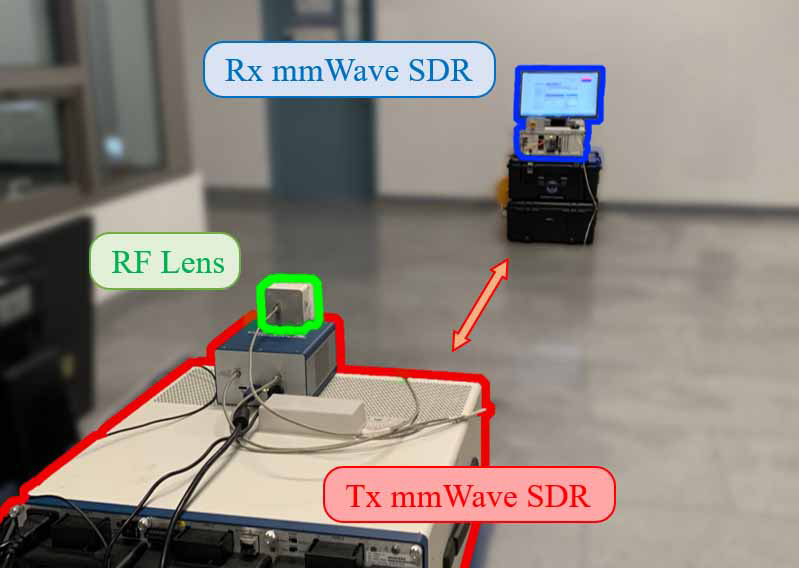}}}}
\hfill
     \subfloat[\label{Fig.lens}]{{\resizebox{0.280\columnwidth}{!}{\includegraphics{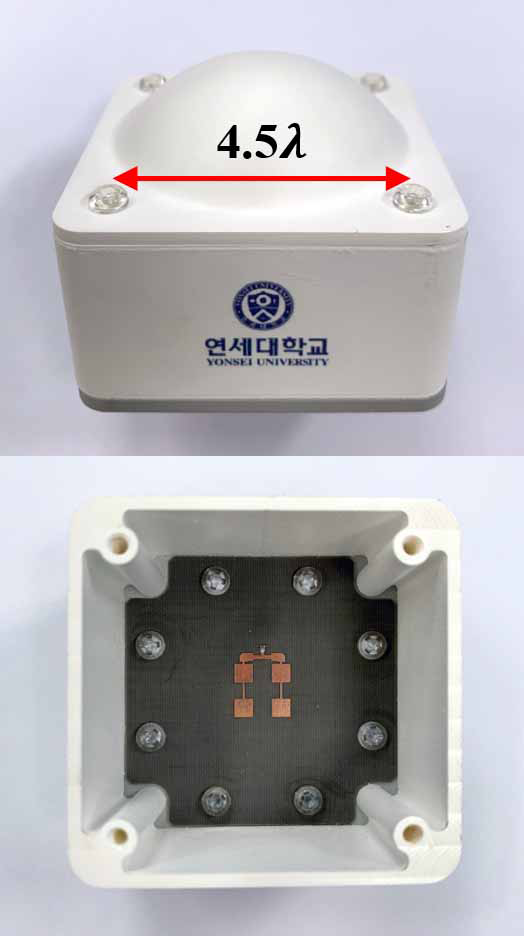}}}}
    \caption{(a) Indoor mmwave software defined radio (SDR) testbed with a fabricated RF lens for beam-squint. (b) fabricated RF lens antenna with polyethenen; below figure is 2 by 2 patch antenna inside lens structure.}
     \label{fig:dummy}
\end{figure}	

Furthermore, the beam-squint phenomena described in the earlier sections are empirically demonstrated to assess performance degradation due to beam-squint in the indoor link-level testbed. In particular, the angle distortion (AD) and power difference (PD) of the lens antenna caused by the wideband, 27.5–29.5 GHz, is evaluated. To assess the BQ phenomenon, we introduce several parameters. First is the AD, which means a difference of azimuth angle of peak power on operating frequency and azimuth angle of peak power on the center frequency (CF). However, It is not possible to judge a hazard only by the angle distortion generated by the beam-squint. The most influential parameters are half power beamwidth (HPBW) and PD. So we need to require the establishment of a new key performance indicator (KPI) as degraded power over beams squint (DPBQ). 

\begin{align}
\begin{split} \label{eq.DPBQ}
&DPBQ~[\%]  \\
&= \frac{(Gain(CF) - PD) \left| \text{sinc} (0.6238~ \frac{AD}{HPBW})\right|}{Gain(CF)} \times 100
\end{split}
\end{align}

 Fig. \ref{Fig.Relation} describes three parameters used in Eq. \ref{eq.DPBQ}. First, PD means the difference between peak antenna gain on each operating frequency and peak antenna gain on the center frequency. Second is the HPBW of the beam on each frequency band. Beamwidth is the main factor that can affect degradation DPBQ and is affected by the array size. Larger array size makes the beam narrower and being critical degradation.  As a result, Table \ref{Link_level} illustrates the AD, PD and DPBQ from beam-squint. 
%빔스퀸트 현상을 평가하기 위해서, 우리는 여러가지 파라미터를 소개한다. 여러파라미터중 가장 잘 받아들여지는 것은 중심주파수 대비 얼마나 각도가 왜곡 되었는지 나타내는 앵글디스토션 (AD) 이다. 일반적으로 빔스퀸트는 AD를 의미한다.    
%Fig. \ref{Fig.Relation} 는 Eq. \ref{eq.SDBQ} 에서 사용하는 3가지 파라미터를 소개한다. 1번 PD는 중심주파수의 빔게인으로 부터 다른 주파수 대역에서 gain차이가 얼마나 나는지를 나타낸다. 2번은 각 주파수대역 마다의 3dB 빔윗스이다. 빔윗스는 안테나 어레이 크기가 커질수록 작아지고, 빔스퀸트관점에서 빔위스가 작을수록(빔이 네로우 할수록) 시스템 성능에 크리티컬하다. 3번 AD는 주파수 대역에 의해서 중심주파수에서 얼마나 각도왜곡이 일어났는지 나타낸다. 본 논문에서, 식 (7)은 3가지 파라미터로 SDBQ를 정의한다. 

%The angle changes were 0.26$^{\circ}$ at 27.5 GHz and 0.08$^{\circ}$ at 29.5 GHz.

\subsection{Link-level Analysis} 
\begin{figure}[t!]
	\centerline{\resizebox{1\columnwidth}{!}{\includegraphics{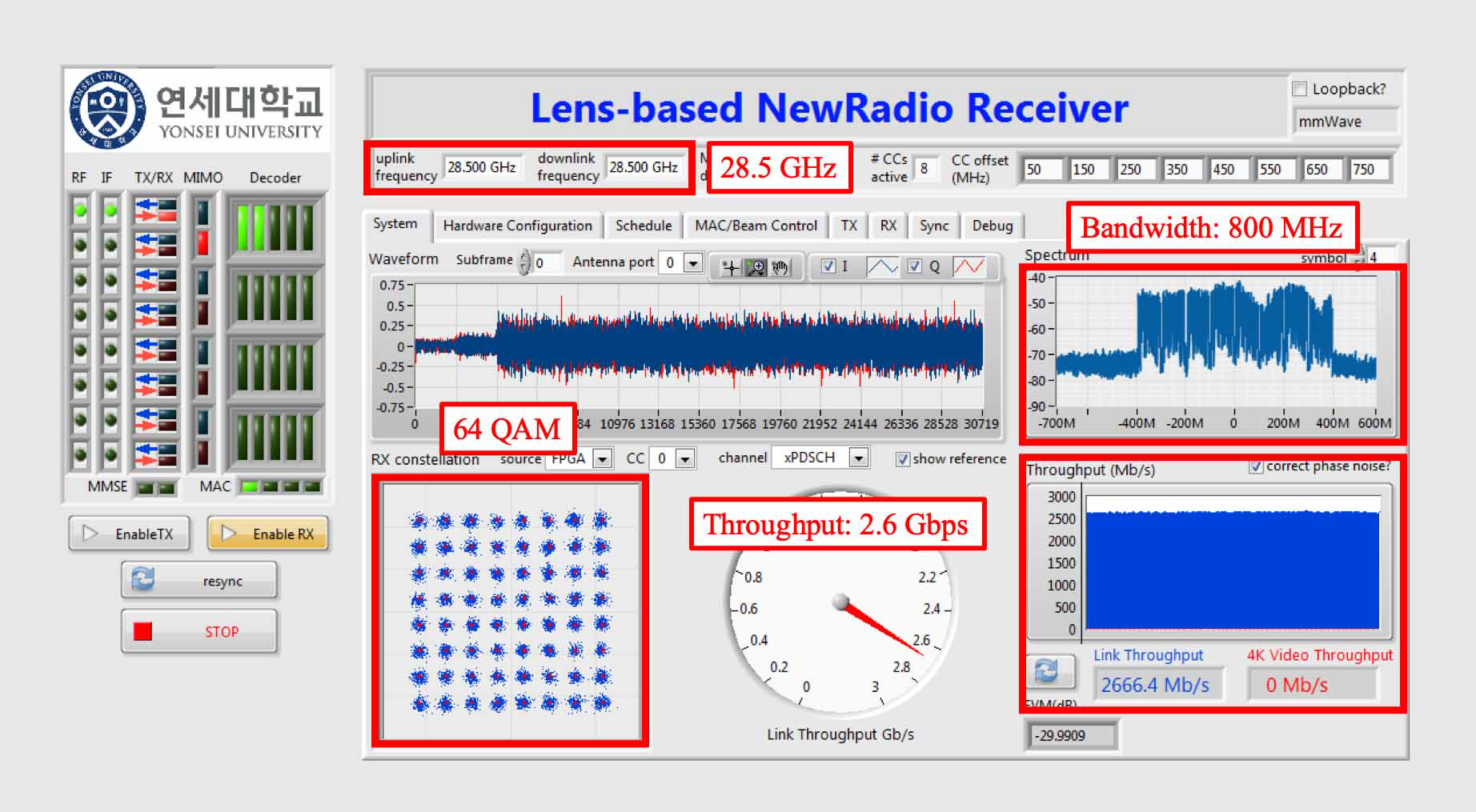}}}
	\caption{Link-level performance analysis at 28.5GHz with 800 MHz bandwidth }\vspace{-1pt}
	\label{Fig.Rx_panel}
\end{figure}

	To test the performance of the lens antenna as a component of a transmission system, in Fig. \ref{Fig.Rx_panel}, the link-level performance is evaluated using the mmWave transceiver system SDR platform with an up/downlink frequency of 28.5 GHz and 800 MHz bandwidth. A single data stream is transmitted and received via the modulation and coding scheme (MCS) (i.e., 64-quadrature amplitude modulation (QAM) and 5/6 coding rate). The results show that the prototyped lens antenna can achieve a maximum throughput of 2.6 Gbps. Note that this high throughput was achieved by a single stream. This is attributed to not only the large bandwidth of the channel but also the high directivity and gain of the lens antenna. Due to this high directivity and gain of the lens antenna, one can deduce that the RF lens-embedded MIMO architecture is a good potential choice for mmWave communication, which has both low cost and good performance.

%We use the PXIe platform and multi-FPGA processing to implement the SDR mmWave transceiver system. 

% % % % % % % % % % % % % % % % % % % % % % % % % % % % % % %
% % % % % % % % % % % % % % % % % % % % % % % % % % % % % % %
% % % % % % % % % % % % % % % % % % % % % % % % % % % % % % %

\begin{table}[t]
\caption{Performance of the fabricated RF lens via mmwave hardware testbed.}
\label{Link_level}
\begin{threeparttable}
{
\begin{tabular}{c|ccccc}
\dtoprule
\textbf{Freqency} & \textbf{azimuth$^{1,*}$} & \textbf{power$^{2,**}$} & \textbf{AD$^{*}$} & \textbf{PD$^{**}$} & \textbf{DPBQ$^{***}$} \\ \midrule
\textbf{27.5 GHz} & 9.52 & -6.9 & -0.26 & -0.6 & 91.98 \\
\textbf{28.5 GHz} & 9.26 & -7.5 & 0 & 0 & 100 \\
\textbf{29.5 GHz} & 9.34 & -6.9 & -0.08 & -0.6 & 91.99 \\
\dbottomrule
\end{tabular}
}
{
        \begin{tablenotes}
        \item[$^{1}$]  appear directivity of analog beamforming in azimuth. 
        \vspace{3pt}
        \item[$^{2}$]  appear received power of the Rx at each azimuth angle. 
        \vspace{5pt}
        \item[$^{*}$, $^{**}$, $^{***}$] Units are [deg], [dBi] and [\%], respectively. 
        \vspace{-5pt}
        \end{tablenotes}
        }
\end{threeparttable}
\end{table}

\section{Conclusion}

% 이 데모스트레이션에서 우리는 lens anten서na를 이용한 mmwave SDR system을 선보인다. 시뮬레이션 분석을 통한 가이드라인을 바탕으로 제작된 렌즈를 통해서,  Lens MIMO communication이 가능함을 보이고 , lens antenna 에서 일어나는 beam-squint 현상이 미미함을 보인다. 

In this paper, we presented the mmWave SDR testbed system with the RF lens-embedded MIMO architecture. Through the fabricated lens based on the guideline from simulation analysis, we verified the feasibility of mmWave lens MIMO communication and showed that the beam-squint phenomenon is insignificant in lens-embedded MIMO architecture.

% use section* for acknowledgment
\section*{Acknowledgment}
This work was supported by Institute of Information \& communications Technology Planning \& Evaluation (IITP) grant funded by the Korea government (MSIP) 
(No. 2016-11-1719), (No. 2018-0-00170).

% trigger a \newpage just before the given reference
% number - used to balance the columns on the last page
% adjust value as needed - may need to be readjusted if
% the document is modified later
% \IEEEtriggeratref{8}
% The "triggered" command can be changed if desired:
%\IEEEtriggercmd{\enlargethispage{-5in}}

% references section
\bibliographystyle{IEEEtran}
%\bibliography{mybibs_MIMO_ML}
	%---------------------------------------------------------------------------------------------%
	%References

% that's all folks
\end{document}